\documentclass[prb,twocolumn,amsmath,amssymb,superscriptaddress,floatfix]{revtex4}

\usepackage{bm}
\usepackage{times}
\usepackage{graphicx}

\begin{document}

\title{Three-Dimensional Topological Insulators in I-III-VI$_2$ and II-IV-V$_2$ Chalcopyrite Semiconductors}

\author{Wanxiang Feng}

\affiliation{Beijing National Laboratory for Condensed Matter Physics and Institute of Physics, Chinese Academy of Sciences, Beijing 100190, China}

\author{Jun Ding}

\affiliation{Beijing National Laboratory for Condensed Matter Physics and Institute of Physics, Chinese Academy of Sciences, Beijing 100190, China}

\author{Di Xiao}

\affiliation{Materials Science \& Technology Division, Oak Ridge National Laboratory, Oak Ridge, TN 37831, USA}

\author{Yugui Yao$^*$}

\affiliation{Beijing National Laboratory for Condensed Matter Physics and Institute of Physics, Chinese Academy of Sciences, Beijing 100190, China}


\maketitle

\textbf{The recent discovery of topological insulators with exotic metallic surface states has garnered great interest in the fields of condensed matter physics and materials science.~\cite{moore2010}  A number of spectacular quantum phenomena have been predicted when the surface states are under the influence of magnetism and superconductivity,~\cite{qi2008,fu2008,tanaka2009,garate2010} which could open up new opportunities for technological applications in spintronics and quantum computing.  To achieve this goal, material realization of topological insulators with desired physical properties is of crucial importance.  Based on first-principles calculations, here we show that a large number of ternary chalcopyrite compounds of composition I-III-VI$_2$ and II-IV-V$_2$ can realize the topological insulating phase in their native states. The crystal structure of chalcopyrites is derived from the frequently used zinc-blende structure, and many of them possess a close lattice match to important mainstream semiconductors, which is essential for a smooth integration into current semiconductor technology.  The diverse optical, electrical and structural properties of chalcopyrite semiconductors,~\cite{shay1975} and particularly their ability to host room-temperature ferromagnetism,~\cite{medvedkin2000,erwin2004} make them appealing candidates for novel spintronic devices.}

Topological insulators are new states of quantum matter characterized by the so-called $Z_2$ topological invariant associated with the bulk electronic structure.~\cite{moore2007,fu2007,roy2009}  Their band structures is usually characterized by a certain type of band-inversion that involves the switching of bands of opposite parity around the Fermi level.~\cite{bernevig2006,fu2007a}  Early material realizations of topological insulators are limited, including the Bi$_{1-x}$Sb$_x$ alloy~\cite{fu2007a,hsieh2008} and the binary tetradymite semiconductors.~\cite{zhang2009,xia2009,chen2009}  Motivated by their application potentials, the search for new topological insulators has moved into ternary compounds as they provide greater material flexibility.~\cite{xiao2010a,lin2010,chadov2010,lin2010a,yan2010,chen2010,sato2010}  For example, it was proposed that in the Heusler family~\cite{xiao2010a,lin2010,chadov2010} the topological insulator can be made with coexisting magnetism, a much desired property for spintronic applications.  However, a clear shortcoming of the Heusler family is that in their native states these materials are either zero-gap semiconductors or semimetals.  Although alloying or proper strain can be used to achieve the insulating behavior, this creates extra complexity in material growth and could introduce detrimental effects upon doping.  In contrast, the chalcopyrite compounds reported in this work have an intrinsic band gap due to its crystal structure, therefore is free from these problems.

\begin{figure}[b]
\includegraphics[width=\columnwidth]{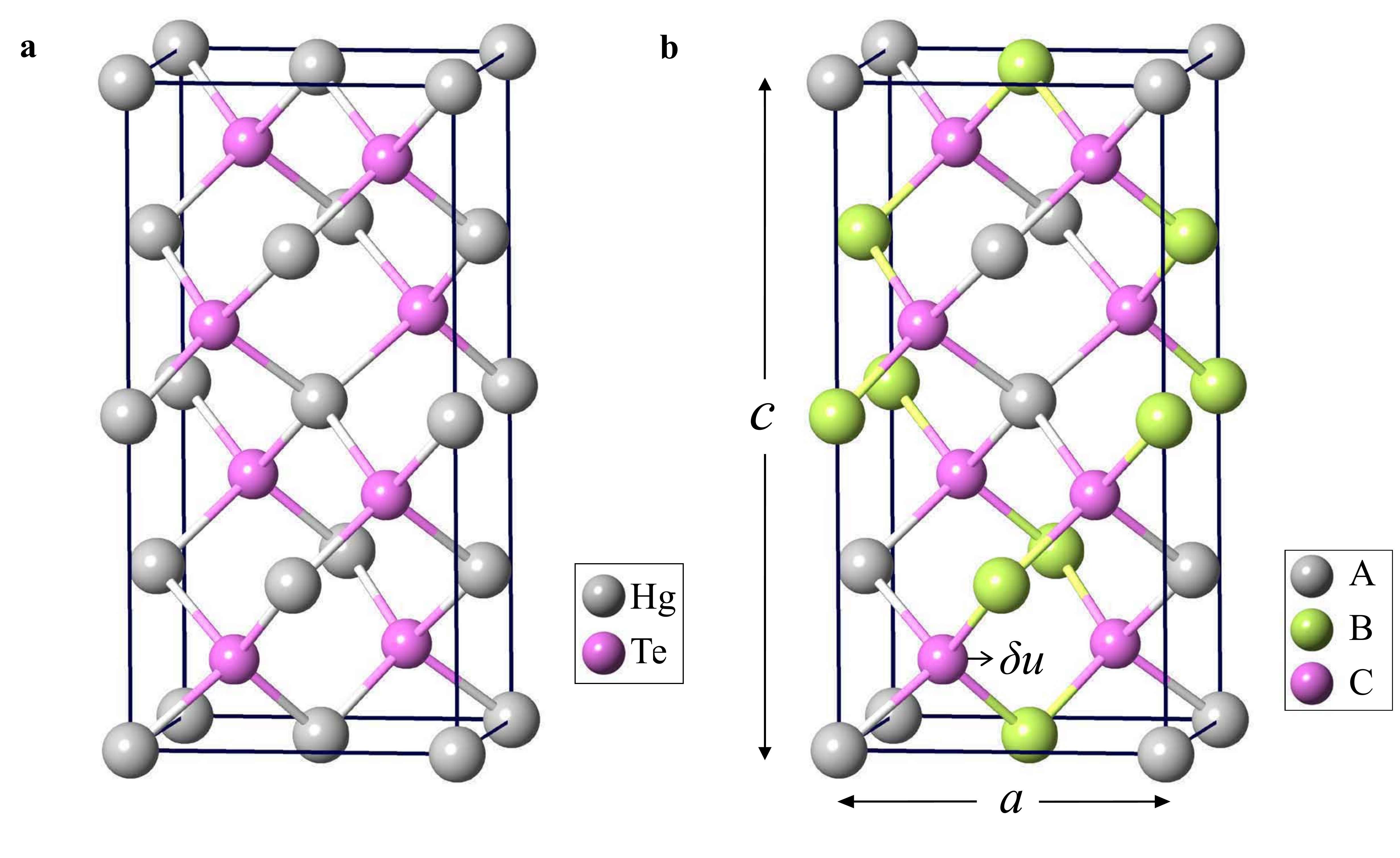}
\caption{\textbf{Comparison of the zinc-blende and the chalcopyrite structure}. \textbf{a}, The zinc-blende compound HgTe.  \textbf{b}, The chalcopyrite compound $ABC_2$.  The internal displacement of the anion is defined as $\delta u = (R_{AC}^2 - R_{BC}^2)/a^2$, where $R_{AC}$ and $R_{BC}$ are the bond lengths between the anion $C$ and its two nearest $A$ and $B$ cations.}
\end{figure}

\begin{figure*}
\includegraphics[width=\textwidth]{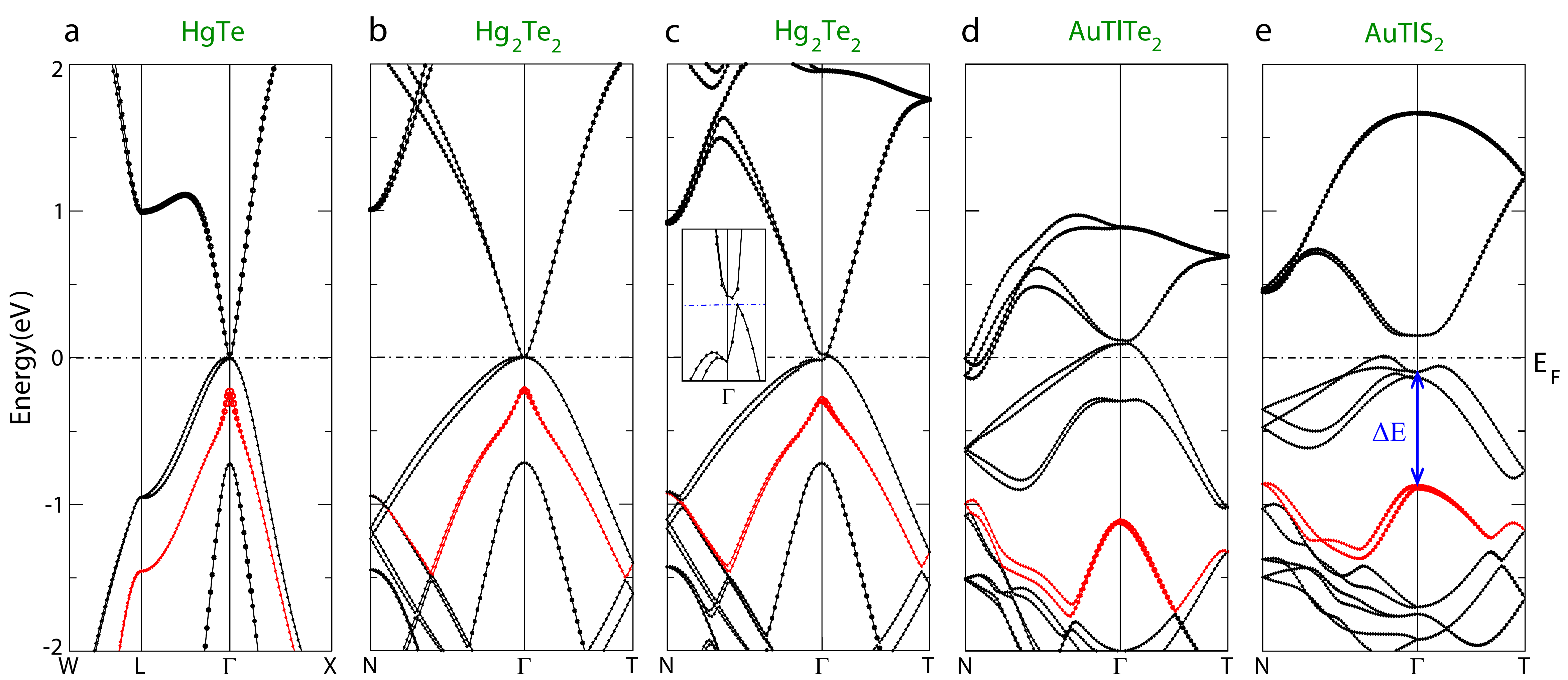}
\caption{\label{fig:band}\textbf{Band structures of HgTe in comparison with AuTlTe$_2$ and AuTlS$_2$.}  \textbf{a-c}, Band structures of HgTe (\textbf{a}), HgTe in an ideal chalcopyrite supercell $(\textbf{b})$, and HgTe in a chalcopyrite cell with $\eta = 1.016$ and $\delta u = -0.018$ (\textbf{c}).  The size of red dots denotes the probability of $s$-orbital occupation on the cations. The insert in \textbf{c} shows the energy gap opened by chalcopyrite distortion. \textbf{d,e}, Band structures of AuTlTe$_2$ (\textbf{d}) and AuTlS$_2$ (\textbf{e}).  Both display an inverted band order similar to HgTe, indicating a nontrivial band topology. $\Delta E$ in \textbf{e} indicates the band inversion strength.}
\end{figure*}

The ternary $ABC_2$ chalcopyrite compounds of composition I-III-VI$_2$ or II-IV-V$_2$ are isoelectronic analogs of the II-VI or III-V binary semiconductors, respectively.~\cite{shay1975}  The crystal lattice of chalcopyrites is described by the space group $D^{12}_{2d}$ ($I\bar{4}2d$), which can be regarded as a superlattice of the zinc-blende structure with small structural distortions.  As illustrated in Fig.~1, the chalcopyrite unit cell is essentially formed by two cubic zinc-blende unit cells, in which the $A$ and $B$ atoms are ordered on the two different cation sites.  In view of the overall structural similarity between the ternary chaclopyrites and their binary analogs, the electronic states of the former are expected to closely resemble those of the latter.  In particular, we anticipate that some of the chalcopyrites might fall into the same topological class as the topologically nontrivial 3D-HgTe,~\cite{fu2007a,bernevig2006} which is a II-VI zinc-blende semiconductor.  Furthermore, the cubic symmetry in chalcopyrite compounds is explicitly broken due to the $AB$ cation ordering, accompanied by two additional structural modifications (Fig.~1b): the tetragonal distortion characterized by the ratio of the lattice constants $\eta = c/2a$, and the internal displacement $\delta u$ of the anions away from their ideal zinc-blende sites.  In an ideal structure, $\eta = 1$ and $\delta u= 0$.  As we show below, this inherent symmetry reduction plays the same role as the uniaxial strain required in cubic topological materials such as 3D-HgTe~\cite{fu2007a} and the Heusler family,~\cite{chadov2010,lin2010,xiao2010a} making it possible to realize the topologically insulating phase in the native state of chalcopyrite compounds.

To illustrate the band topology of the ternary chalcopyrites, we begin with their binary analog, HgTe, and follow its band structure under an adiabatic transformation during which the zinc-blende structure is gradually evolved into the chalcopyrite structure (Fig.~\ref{fig:band}a-c).  The first-principles relativistic band structure is obtained by using the full-potential linearized augmented plane-wave method implemented in the \textsc{wien2k} package.~\cite{wien2k}  The exchange-correlation potential is treated using the modified Becke-Johnson potential,~\cite{becke2006} which is a semi-local potential designed to accurately predict the band gap.~\cite{tran2009}  The band structure of HgTe features a distinctive inverted band order at the $\Gamma$ point, where the $s$-like $\Gamma_6$ states sit below the four-fold degenerate $p$-like $\Gamma_8$ states (Fig.~\ref{fig:band}a).  Away from the $\Gamma$ point, the valence and conduction bands are well separated without crossing each other.  Since the band inversion occurs only once throughout the Brillouin zone, HgTe is in a topologically nontrivial phase.~\cite{fu2007a}  The four-fold degeneracy of the $\Gamma_8$ states is protected by the cubic symmetry of the zinc-blende structure, thus in its native state HgTe is a zero-gap semiconductor.  The band structure calculation is then repeated for HgTe in an ideal chalcopyrite supercell (Fig.~\ref{fig:band}b), which is four times larger than the zinc-blende one.  During this procedure, every set of four different wavevectors of the original zinc-blende Brillouin zone fold into a single point of the chalcopyrite Brillouin zone.  In particular, states at the original $\Gamma$ point fold back to themselves, therefore the low-energy electronic properties are still dominated by states around the $\Gamma$ point.  In the final step, to mimic the cubic-symmetry breaking in chalcopyrites, we let the system undergo a tetragonal distortion plus an internal displacement of the two Te atoms.  This structural change lifts the degeneracy of the $\Gamma_8$ states and opens up an energy gap, while leaving the inverted band order intact (Fig.~\ref{fig:band}c).  The situation here thus is exactly like HgTe under a uniaxial strain,~\cite{fu2007a,bernevig2006} and the resulting compound, an artificial chalcopyrite Hg$_2$Te$_2$, is a topological insulator.

\begin{figure*}
\includegraphics[width=12cm]{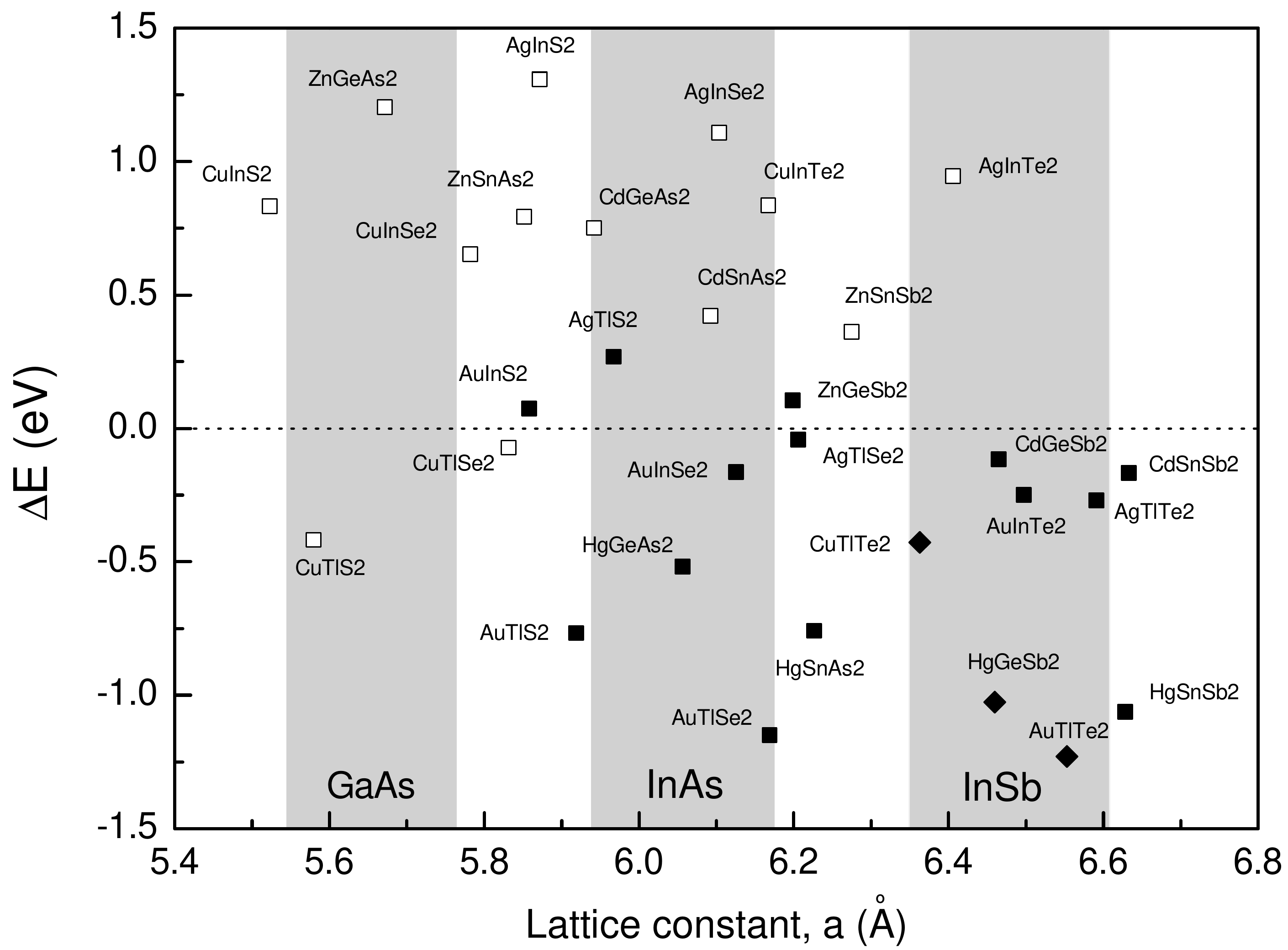}
\caption{\label{fig:map}\textbf{Inverted band strength for various chalcopyrites as a function of the lattice constant.}  Open symbols mark the compounds whose lattice constant was reported in the literature.  For the rest of the compounds their equilibrium lattice constants are obtained by first-principles total energy minimization.  When $\Delta E < 0$, squares mark the compounds that are topological insulators and diamonds mark the systems that are topological metals.  Shaded areas indicate materials that are expected to be closely $(\pm 2\%)$ lattice matched to either GaAs, InAs, or InSb.}
\end{figure*}

The above analysis of the tetragonally distorted HgTe suggests that the same $Z_2$ band topology can be realized in real ternary chalcopyrites.  This is confirmed by our extensive first-principles band structure calculations over this family of materials.  There are two possible scenarios, with the corresponding band structures of representative systems shown in Fig.~\ref{fig:band}d and 2e.  The most natural generalization of HgTe to chalcopyrites is the ternary compound AuTlTe$_2$, which can be obtained by replacing the two Hg atoms in Hg$_2$Te$_2$ with an isoelectronic pair of Au and Tl atoms.  Conceptually, this process can be described by an adiabatic transformation in which the nuclear charges on the two Hg sites are continuously changed to that of Au and Tl, respectively.  However, such a substitution causes the conduction bands to drop below the Fermi level at the $N$ point, making the system effectively a semimetal.  Note that a $Z_2$ topological invariant can still be defined for the valence bands in AuTlTe$_2$ because they are separated from the conduction bands by local energy gap throughout the Brillouin zone.  Our further study shows that the semimetallic behavior can be removed by choosing a lighter element S in place of Te---we find that a global band gap of 0.14 eV can be realized in AuTlS$_2$ (Fig.~\ref{fig:band}e).  Remarkably, this band gap is much larger than typical band gaps opened by uniaxial strain in Heusler compounds,~\cite{chadov2010,lin2010,xiao2010a} and is also well above the energy scale of room temperature.  In both AuTlTe$_2$ and AuTlS$_2$, the inverted band order at the $\Gamma$ point is retained, with no band inversion at other places.  We therefore conclude that, in their native states, AuTlS$_2$ is a topological insulator with a sizable band gap, and AuTlTe$_2$ a topological metal.    On the other hand, compounds with normal band order are always semiconductors---our calculation did not find any semimetallic chalcopyrite with normal band order.

\begin{figure}[b]
\includegraphics[width=8cm]{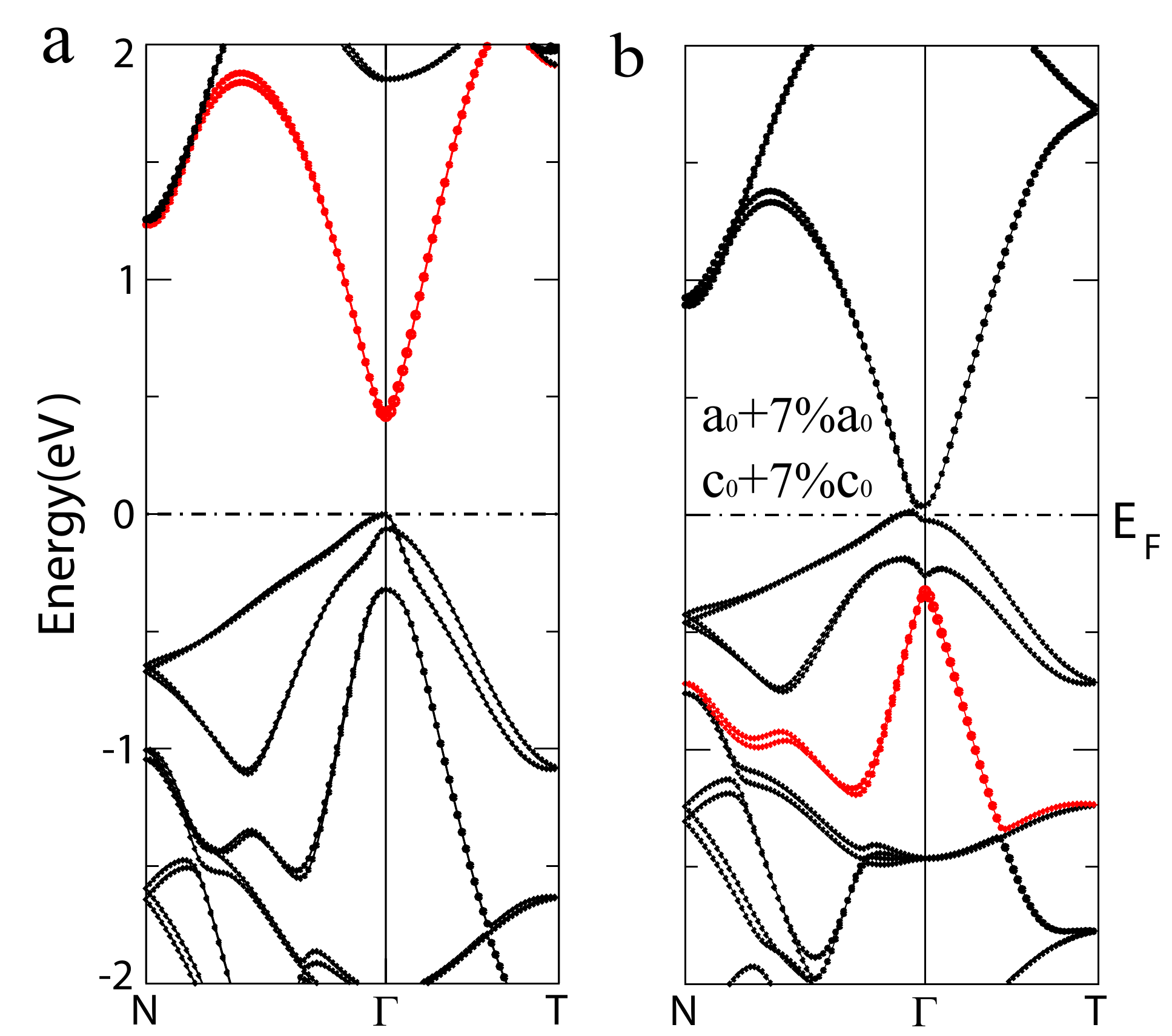}
\caption{\label{fig:convert}\textbf{Band structures of CdSnAs$_2$ without and with a hydrostatic strain.}  The application of a hydrostatic strain causes the $s$-like $\Gamma_6$ bands (marked by red dots) to jump below the valence band top, providing the necessary band inversion that leads to a nontrivial topological order.}
\end{figure}

So far our argument has been based on the isoelectronic relation between the ternary chalcopyrites and their binary analogs; it is much desired to give a definite proof that the band topology of AuTlTe$_2$ and AuTlS$_2$ is $Z_2$ nontrivial.  Because of the lack of inversion symmetry in these materials, the parity criterion developed by Fu and Kane~\cite{fu2007a} does not apply.  Here we directly evaluate the $Z_2$ invariant in terms of the Berry gauge potential and Berry curvature associated with the Bloch functions.~\cite{xiao2010}  In this approach, the $Z_2$ invariant is interpreted as an obstruction to smoothly defining the wave function throughout the Brillouin zone under a gauge that respects the time-reversal symmetry.~\cite{moore2007,fu2006}  Unlike the parity criterion, this approach does not require any specific point-group symmetry.  Following ref.~\onlinecite{fukui2007}, we have carried out the lattice computation of the $Z_2$ invariant based on our first-principles method (see ref.~\onlinecite{xiao2010a} and Supplementary Information for details).  We find that, for all the compounds presented in this work, the calculated $Z_2$ invariant agrees with the intuitive band-inversion picture. In practice, this translates into an empirical rule that if the $s$-orbital originated $\Gamma_6$ states are completely occupied and below the valence band maximum, the compound possesses a nontrivial topological band structure.

Armed with the above insight, next we explore the band topology of the chalcopyrite family of the I-III-VI$_2$ compounds (I = Cu, Ag, Au; III = In, Tl; V = S, Se, Te), as well as the II-IV-V$_2$ compounds (II = Zn, Cd, Hg; IV = Ge, Sn; V = As, Sb).  Since the band inversion only occurs at the $\Gamma$ point, the band order at that point can be used to characterize the band topology.  In compounds with cubic symmetry, one can define the band inversion strength as the energy difference between the $\Gamma_6$ and $\Gamma_8$ states.~\cite{lin2010,chadov2010} However, in passing from the zinc-blende structure to the chalcopyrite structure, the original $\Gamma_8$ states split into states with $\Gamma_7$ and $\Gamma_6$ symmetry, which typically form the top set of the valence bands and the bottom set of the conduction bands.  We therefore define the band inversion strength $\Delta E$ as the energy difference between the $s$-orbital originated $\Gamma_6$ states and the valence band maximum at the $\Gamma$ point.  Figure~\ref{fig:map} shows $\Delta E$ for a number of chalcopyrite compounds as a function of the lattice constant.  Materials with $\Delta E > 0$ are topologically trivial and those with $\Delta E < 0$ are either topological insulators or topological metals.  As shown in Fig.~\ref{fig:map}, there are quite a few chalcopyrite topological insulators with a close lattice matching to several mainstream semiconductors.  Finally, we note that the impressive collection of chalcopyrite topological insulators can be further extended by proper strain engineering.  Similar to the Heusler family,~\cite{chadov2010,lin2010,xiao2010a} the topological phase is sensitive to the lattice constant, which offers certain degree of tunability of these materials.  For example, a 7\% change in the lattice constant converts the trivial semiconductor CdSnAs$_2$ into a topological insulator (Fig.~\ref{fig:convert}).  

The interplay between topological surface states and ferromagnetism has been under extensive investigation because of the possibility of electric control of magnetization with low power consumption.~\cite{qi2008,garate2010}  The wide availability of chalcopyrite semiconductors, the control of topological order via lattice tuning, and their excellent prospect for room temperature ferromagnetism, make them the ideal platform to further investigate these novel topological phenomena and could enable the next-generation spintronic devices.

\textbf{Acknowledgement.} We acknowledges useful discussions with Jun Wen, Wenguang Zhu and Hanno Weitering.  Y.G.Y acknowledges support from the NSF of China (10674163, 10974231), the MOST Project of China (2006CB921300, 2007CB925000), and Supercomputing Center of Chinese Academy of Sciences.  D.X is supported by the Division of Materials Sciences and Engineering, Office of Basic Energy Sciences, U.S. Department of Energy.

\textbf{Author contributions.} Y.G.Y. and D.X. conceived the idea of searching for topological insulators in ternary chalcopyrites and supervised the overall project.  W.F. carried out the main part of the calculation with assistance from Y.G.Y., J.D. also helped with the calculation.

\textbf{Additional information.} The authors declare no competing financial interests. Reprints and permissions information is available online at http://npg.nature.com/reprintsandpermissions. Correspondence and requests for materials should be addressed to Y.G.Y. (ygyao@aphy.iphy.ac.cn)


\begin{thebibliography}{10}
\expandafter\ifx\csname url\endcsname\relax
  \def\url#1{\texttt{#1}}\fi
\expandafter\ifx\csname urlprefix\endcsname\relax\def\urlprefix{URL }\fi
\providecommand{\bibinfo}[2]{#2}
\providecommand{\eprint}[2][]{\url{#2}}

\bibitem{moore2010}
\bibinfo{author}{Moore, J.~E.}
\newblock \bibinfo{title}{The birth of topological insulators}.
\newblock \emph{\bibinfo{journal}{Nature}} \textbf{\bibinfo{volume}{464}},
  \bibinfo{pages}{194--198} (\bibinfo{year}{2010}).

\bibitem{qi2008}
\bibinfo{author}{Qi, X.-L.}, \bibinfo{author}{Hughes, T.~L.} \&
  \bibinfo{author}{Zhang, S.-C.}
\newblock \bibinfo{title}{Topological field theory of time-reversal invariant
  insulators}.
\newblock \emph{\bibinfo{journal}{Phys. Rev. B}} \textbf{\bibinfo{volume}{78}},
  \bibinfo{pages}{195424} (\bibinfo{year}{2008}).

\bibitem{fu2008}
\bibinfo{author}{Fu, L.} \& \bibinfo{author}{Kane, C.~L.}
\newblock \bibinfo{title}{Superconducting proximity effect and majorana
  fermions at the surface of a topological insulator}.
\newblock \emph{\bibinfo{journal}{Phys. Rev. Lett.}}
  \textbf{\bibinfo{volume}{100}}, \bibinfo{pages}{096407}
  (\bibinfo{year}{2008}).

\bibitem{tanaka2009}
\bibinfo{author}{Tanaka, Y.}, \bibinfo{author}{Yokoyama, T.} \&
  \bibinfo{author}{Nagaosa, N.}
\newblock \bibinfo{title}{Manipulation of the {M}ajorana fermion, {A}ndreev
  reflection, and {J}osephson current on topological insulators}.
\newblock \emph{\bibinfo{journal}{Phys. Rev. Lett.}}
  \textbf{\bibinfo{volume}{103}}, \bibinfo{pages}{107002}
  (\bibinfo{year}{2009}).

\bibitem{garate2010}
\bibinfo{author}{Garate, I.} \& \bibinfo{author}{Franz, M.}
\newblock \bibinfo{title}{Inverse spin-galvanic effect in the interface between
  a topological insulator and a ferromagnet}.
\newblock \emph{\bibinfo{journal}{Phys. Rev. Lett.}}
  \textbf{\bibinfo{volume}{104}}, \bibinfo{pages}{146802}
  (\bibinfo{year}{2010}).

\bibitem{shay1975}
\bibinfo{author}{Shay, J.~L.} \& \bibinfo{author}{Wernick, J.~H.}
\newblock \emph{\bibinfo{title}{Ternary Chalcopyrite Semiconductors: Growth,
  Electronic Properties and Applications}} (\bibinfo{publisher}{Pergamon
  Press}, \bibinfo{address}{Oxford}, \bibinfo{year}{1975}).

\bibitem{medvedkin2000}
\bibinfo{author}{Medvedkin, G.~A.} \emph{et~al.}
\newblock \bibinfo{title}{Room temperature ferromagnetism in novel diluted
  magnetic semiconductor {Cd}$_{1-x}${Mn}$_{x}${Ge}{P}$_{2}$}.
\newblock \emph{\bibinfo{journal}{Jpn. J. Appl. Phys.}}
  \textbf{\bibinfo{volume}{39}}, \bibinfo{pages}{L949--L951}
  (\bibinfo{year}{2000}).
\bibinfo{author}{Cho, S.} \emph{et~al.}
\newblock \bibinfo{title}{Room-temperature ferromagnetism in
  ({Zn}$_{1-x}${Mn}$_x$){Ge}{P}$_2$ semiconductors}.
\newblock \emph{\bibinfo{journal}{Phys. Rev. Lett.}}
  \textbf{\bibinfo{volume}{88}}, \bibinfo{pages}{257203}
  (\bibinfo{year}{2002}).

\bibitem{erwin2004}
\bibinfo{author}{Erwin, S.~C.} \& \bibinfo{author}{Zutic, I.}
\newblock \bibinfo{title}{Tailoring ferromagnetic chalcopyrites}.
\newblock \emph{\bibinfo{journal}{Nature Mater.}} \textbf{\bibinfo{volume}{3}},
  \bibinfo{pages}{410--414} (\bibinfo{year}{2004}).

\bibitem{moore2007}
\bibinfo{author}{Moore, J.~E.} \& \bibinfo{author}{Balents, L.}
\newblock \bibinfo{title}{Topological invariants of time-reversal-invariant
  band structures}.
\newblock \emph{\bibinfo{journal}{Phys. Rev. B}} \textbf{\bibinfo{volume}{75}},
  \bibinfo{pages}{121306} (\bibinfo{year}{2007}).

\bibitem{fu2007}
\bibinfo{author}{Fu, L.}, \bibinfo{author}{Kane, C.~L.} \&
  \bibinfo{author}{Mele, E.~J.}
\newblock \bibinfo{title}{Topological insulators in three dimensions}.
\newblock \emph{\bibinfo{journal}{Phys. Rev. Lett.}}
  \textbf{\bibinfo{volume}{98}}, \bibinfo{pages}{106803}
  (\bibinfo{year}{2007}).

\bibitem{roy2009}
\bibinfo{author}{Roy, R.}
\newblock \bibinfo{title}{Topological phases and the quantum spin Hall effect
  in three dimensions}.
\newblock \emph{\bibinfo{journal}{Phys. Rev. B}} \textbf{\bibinfo{volume}{79}},
  \bibinfo{pages}{195322} (\bibinfo{year}{2009}).

\bibitem{bernevig2006}
\bibinfo{author}{Bernevig, B.~A.}, \bibinfo{author}{Hughes, T.~L.} \&
  \bibinfo{author}{Zhang, S.-C.}
\newblock \bibinfo{title}{Quantum spin {H}all effect and topological phase
  transition in {HgTe} quantum wells}.
\newblock \emph{\bibinfo{journal}{Science}} \textbf{\bibinfo{volume}{314}},
  \bibinfo{pages}{1757--1761} (\bibinfo{year}{2006}).

\bibitem{fu2007a}
\bibinfo{author}{Fu, L.} \& \bibinfo{author}{Kane, C.~L.}
\newblock \bibinfo{title}{Topological insulators with inversion symmetry}.
\newblock \emph{\bibinfo{journal}{Phys. Rev. B}} \textbf{\bibinfo{volume}{76}},
  \bibinfo{pages}{045302} (\bibinfo{year}{2007}).

\bibitem{hsieh2008}
\bibinfo{author}{Hsieh, D.} \emph{et~al.}
\newblock \bibinfo{title}{A topological Dirac insulator in a quantum spin hall
  phase}.
\newblock \emph{\bibinfo{journal}{Nature}} \textbf{\bibinfo{volume}{452}},
  \bibinfo{pages}{970--974} (\bibinfo{year}{2008}).

\bibitem{zhang2009}
\bibinfo{author}{Zhang, H.} \emph{et~al.}
\newblock \bibinfo{title}{Topological insulators in {Bi}$_2${Se}$_3$,
  {Bi}$_2${Te}$_3$ and {Sb}$_2${Te}$_3$ with a single {D}irac cone on the
  surface}.
\newblock \emph{\bibinfo{journal}{Nature Phys.}} \textbf{\bibinfo{volume}{5}},
  \bibinfo{pages}{438--442} (\bibinfo{year}{2009}).

\bibitem{xia2009}
\bibinfo{author}{Xia, Y.} \emph{et~al.}
\newblock \bibinfo{title}{Observation of a large-gap topological-insulator
  class with a single Dirac cone on the surface}.
\newblock \emph{\bibinfo{journal}{Nature Phys.}} \textbf{\bibinfo{volume}{5}},
  \bibinfo{pages}{398--402} (\bibinfo{year}{2009}).

\bibitem{chen2009}
\bibinfo{author}{Chen, Y.~L.} \emph{et~al.}
\newblock \bibinfo{title}{Experimental realization of a three-dimensional
  topological insulator, {Bi}$_2${Te}$_3$}.
\newblock \emph{\bibinfo{journal}{Science}} \textbf{\bibinfo{volume}{325}},
  \bibinfo{pages}{178--181} (\bibinfo{year}{2009}).

\bibitem{xiao2010a}
\bibinfo{author}{Xiao, D.},
\bibinfo{author}{Yao, Y.G.},
\bibinfo{author}{Feng, W.X.}
 \emph{et~al.}
\newblock \bibinfo{title}{Half-{H}eusler compounds as a new class of
  three-dimensional topological insulators}.
\newblock \emph{\bibinfo{journal}{Phys. Rev. Lett.}} (in press).

\bibitem{lin2010}
\bibinfo{author}{Lin, H.} \emph{et~al.}
\newblock \bibinfo{title}{Half-{H}eusler ternary compounds as new
  multifunctional experimental platforms for topological quantum phenomena}.
\newblock \emph{\bibinfo{journal}{Nature Mater.}} \textbf{\bibinfo{volume}{9}},
  \bibinfo{pages}{546--549} (\bibinfo{year}{2010}).

\bibitem{chadov2010}
\bibinfo{author}{Chadov, S.} \emph{et~al.}
\newblock \bibinfo{title}{Tunable multifunctional topological insulators in
  ternary {H}eusler compounds}.
\newblock \emph{\bibinfo{journal}{Nature Mater.}} \textbf{\bibinfo{volume}{9}},
  \bibinfo{pages}{541--545} (\bibinfo{year}{2010}).

\bibitem{lin2010a}
\bibinfo{author}{Lin, H.} \emph{et~al.}
\newblock \bibinfo{title}{Single-Dirac-cone topological surface states in the
  {TlBiSe}$_2$ class of topological semiconductors}.
\newblock \emph{\bibinfo{journal}{Phys. Rev. Lett.}}
  \textbf{\bibinfo{volume}{105}}, \bibinfo{pages}{036404}
  (\bibinfo{year}{2010}).

\bibitem{yan2010}
\bibinfo{author}{Yan, B.} \emph{et~al.}
\newblock \bibinfo{title}{Theoretical prediction of topological insulators in
  thallium-based III-V-VI$_2$ ternary chalcogenides}.
\newblock \emph{\bibinfo{journal}{Europhys. Lett.}}
  \textbf{\bibinfo{volume}{90}}, \bibinfo{pages}{37002} (\bibinfo{year}{2010}).

\bibitem{chen2010}
\bibinfo{author}{Chen, Y.} \& \bibinfo{author}{Others}.
\newblock \bibinfo{title}{Observation of single Dirac cone topological surface
  state in compounds {Tl}{Bi}{Te}$_2$ and {Tl}{Bi}{Se}$_2$ from a new
  topological insulator family}.
\newblock \eprint{arXiv:1006.3843}.

\bibitem{sato2010}
\bibinfo{author}{Sato, T.} \& \bibinfo{author}{Others}.
\newblock \bibinfo{title}{Direct evidence for the Dirac-cone topological
  surface states in ternary chalcogenide {Tl}{Bi}{Se}$_2$}.
\newblock \eprint{arXiv:1006.2437}.

\bibitem{wien2k}
\bibinfo{author}{Blaha, P.}, \bibinfo{author}{Schwarz, K.},
  \bibinfo{author}{Madsen, G. K.~H.}, \bibinfo{author}{Kvasnicka, D.} \&
  \bibinfo{author}{Luitz, J.}
\newblock \emph{\bibinfo{title}{\textsc{wien2k}: An Augmented Plane Wave and
  Local Orbitals Program for Calculating Crystal Properties}}
  (\bibinfo{publisher}{Vienna University of Technology}, \bibinfo{year}{2001}).

\bibitem{becke2006}
\bibinfo{author}{Becke, A.~D.} \& \bibinfo{author}{Johnson, E.~R.}
\newblock \bibinfo{title}{A simple effective potential for exchange}.
\newblock \emph{\bibinfo{journal}{J. Chem. Phys.}}
  \textbf{\bibinfo{volume}{124}}, \bibinfo{pages}{221101}
  (\bibinfo{year}{2006}).

\bibitem{tran2009}
\bibinfo{author}{Tran, F.} \& \bibinfo{author}{Blaha, P.}
\newblock \bibinfo{title}{Accurate band gaps of semiconductors and insulators
  with a semilocal exchange-correlation potential}.
\newblock \emph{\bibinfo{journal}{Phys. Rev. Lett.}}
  \textbf{\bibinfo{volume}{102}}, \bibinfo{pages}{226401}
  (\bibinfo{year}{2009}).

\bibitem{xiao2010}
\bibinfo{author}{Xiao, D.}, \bibinfo{author}{Chang, M.-C.} \&
  \bibinfo{author}{Niu, Q.}
\newblock \bibinfo{title}{Berry phase effects on electronic properties}.
\newblock \emph{\bibinfo{journal}{Rev. Mod. Phys.}}
  \textbf{\bibinfo{volume}{82}}, \bibinfo{pages}{1959--2007}
  (\bibinfo{year}{2010}).

\bibitem{fu2006}
\bibinfo{author}{Fu, L.} \& \bibinfo{author}{Kane, C.~L.}
\newblock \bibinfo{title}{Time reversal polarization and a {Z}$_2$ adiabatic
  spin pump}.
\newblock \emph{\bibinfo{journal}{Phys. Rev. B}} \textbf{\bibinfo{volume}{74}},
  \bibinfo{pages}{195312} (\bibinfo{year}{2006}).

\bibitem{fukui2007}
\bibinfo{author}{Fukui, T.} \& \bibinfo{author}{Hatsugai, Y.}
\newblock \bibinfo{title}{Quantum spin {H}all effect in three dimensional
  materials: Lattice computation of {Z}$_2$ topological invariants and its
  application to {Bi} and {Sb}}.
\newblock \emph{\bibinfo{journal}{J. Phys. Soc. Jpn.}}
  \textbf{\bibinfo{volume}{76}}, \bibinfo{pages}{053702}
  (\bibinfo{year}{2007}).


\end{thebibliography}
\end{document}